\documentclass{article}

\usepackage{PRIMEarxiv}

\usepackage[utf8]{inputenc} 
\usepackage[T1]{fontenc}    
\usepackage{hyperref}       
\usepackage{url}            
\usepackage{booktabs}       
\usepackage{amsfonts}       
\usepackage{nicefrac}       
\usepackage{microtype}      
\usepackage{lipsum}
\usepackage{fancyhdr}       
\usepackage{graphicx}       
\usepackage{algorithm}
\usepackage{algpseudocode}
\usepackage{amsmath}
\usepackage{caption}
\graphicspath{{media/}}     

\title{Stain Normalization of Hematology Slides using Neural Color Transfer
}

\author{
  M. Muneeb Arshad \\
  School of Mechanical and Manufacturing Engineering \\
  National University of Sciences and Technology \\
  Islamabad, Pakistan\\
  \texttt{14bememmarshad@smme.edu.pk}
   \And
  Hasan Sajid \\
  School of Mechanical and Manufacturing Engineering \\
  National University of Sciences and Technology \\
  Islamabad, Pakistan\\
  \texttt{hasan.sajid@smme.nust.edu.pk}
  \And
  M. Jawad Khan \\
  School of Mechanical and Manufacturing Engineering \\
  National University of Sciences and Technology \\
  Islamabad, Pakistan\\
  \texttt{jawad.khan@smme.nust.edu.pk}
}

\begin{document}
\maketitle

\begin{abstract}
Deep learning is popularly used for analyzing pathology images, but variations in image properties can limit the effectiveness of the models. The study aims to develop a method that transfers the variability present in the training set to unseen images, improving the model's ability to make accurate inferences. YOLOv5 was trained on peripheral blood and bone marrow sample images and Neural Color Transfer techniques were used to incorporate invariance. The results showed significant improvement in detecting WBCs from untrained samples after normalization, highlighting the potential of deep learning-based normalization techniques for inference robustness.
\end{abstract}

\keywords{Deep learning in pathology \and Pathology image analysis \and Variance transfer \and Robust inference models \and Image acquisition variations \and Unseen image properties \and Accurate model inferences}

\section{Introduction}
Accurate counting and classification of white blood cells (WBCs) are critical for diagnosing various diseases. Two common methods for counting WBCs are hematology analyzers and manual counting, while cell morphology testing helps diagnose leukemia by classifying cells based on their size and shape, where a lab specialist examines air-dried peripheral blood or bone marrow aspirate stained with a dye under a microscope. Abnormalities and immature white blood cells can be detected by analyzing the percentage of WBCs in the blood. Tremendous research has been conducted to develop alternative methods for WBC classification and counting using image processing techniques. However, the accuracy of these methods decreases when the image properties vary due to external factors such as equipment noise and staining inconsistencies.

Stain normalization is a technique used in digital pathology and hematology to reduce color variation in stained tissue images. The process involves transforming the color appearance of images to make them appear more uniform so that computer algorithms can better detect features and make more accurate diagnoses. This is important because different laboratory procedures or staining protocols can cause significant variability in the appearance of tissue samples, leading to inconsistent results. Stain normalization methods involve statistical analysis of the color channels in a set of images, followed by the transformation of these channels to align the color distribution across images.

Commonly, conventional methods for stain normalization include stain separation and color-matching methods. Color matching methods try to match the color distribution of the source image to that of a reference image \cite{shaban2019staingan}. For example, Reinhard et al. \cite{reinhard2001color} presented a method for matching the colorspace distribution of one image to that of another image by using LAB colorspace so as to match the standard deviation and mean of each color channel in the two images relative to that colorspace. On the other hand, stain-separation methods separate and normalize each staining channel of the source image
independently. Macenko et al. \cite{macenko2009method} and Vahadane et al. \cite{vahadane2016structure} respectively used mathematical methods to compute stain vectors using singular value decomposition (SVD) in Optical Density (OD) space, and sparse non-negative matrix factorization (SNMF). Nevertheless, almost all of these methods depend on a reference image to assess stain parameters. However, it is difficult to cover all staining peculiarities from one reference image, which usually causes inaccurate stain parameters and consequently delivers incorrect color transfer.

Generative Adversarial Networks (GANs) are deep learning-based methods commonly used for stain normalization \cite{cai2019stain,salehi2020pix2pix,shaban2019staingan, zhu2017unpaired,tellez2019quantifying, cho2017neural}. Shaban et al. \cite{shaban2019staingan} proposed a CycleGAN \cite{zhu2017unpaired} based unsupervised stain normalization technique called StainGAN. Cai et al. \cite{cai2019stain} introduced a new generator for obtaining better image quality and speeding up the network performance. Shaojin et al. \cite{cho2017neural}, Salehi et al. \cite{salehi2020pix2pix}, and Tellez et al. \cite{tellez2019quantifying} recreated original images from color-augmented images. However, it is hard to maintain the source image
properties due to the instability of GANs and the complexities of deep neural networks as they
can introduce noise and structural distortions which can have adverse effects on pathological diagnosis \cite{lei2020staincnns}. Nevertheless, deep learning networks usually contain around a million parameters, so their computational cost is high \cite{zheng2020stain}.

This study offers a solution to the aforementioned challenges by comparing the results of WBC detection and classification before and after normalization. The findings indicate that normalization enhances accuracy, thereby presenting significant benefits for patient care and medical research.

\section{Methodology}\label{Methodology}
\subsection{Data Gathering}\label{Dataset}
\begin{figure}[h]
\centering
\includegraphics[width=1\textwidth]{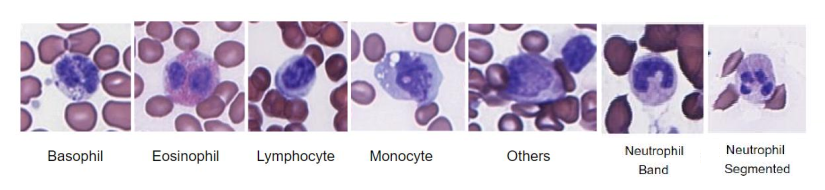}
\caption{Cell types in the private dataset}\label{fig2}
\label{fig:classes}
\end{figure}

The dataset used in this study consists of 7 classes of White Blood Cells (WBCs) with distinctive characteristics as shown in Fig \ref{fig:classes}. Neutrophils are characterized by their pink/orange cytoplasm. Neutrophil Segmented has segmented nuclei, while Neutrophil Band has a single-lobed nucleus that may appear pinched. Basophils have large granules that obscure the nucleus, while Eosinophils have a reddish cytoplasm and bi-lobed nuclei. Lymphocytes have a large dark-staining nucleus with little cytoplasm, and Monocytes have an oval nucleus with a high nucleus-to-cytoplasm ratio. The Others class encompasses all other complicated cell classes or debris. The equipment used for image acquisition was Leica DM2000 Microscope under 40x zoom resolution.  A total of 3226 images were used in the training set, while 807 images were used in the test set. 

\begin{table}[h]
\centering
\begin{minipage}{174pt}
\caption{Train Set Class Distribution}\label{tab1}%
\begin{tabular}{@{}ll@{}}
\toprule
Class & Count \\
\midrule
Neutrophil Segmented    & 22294 \\
Lymphocyte    & 12737 \\
Monocyte   & 4756 \\
Neutrophil Band & 3509 \\
Others & 2167 \\
Eosinophil & 1465 \\
Basophil & 264 \\
\bottomrule
\end{tabular}
\end{minipage}
\begin{minipage}{174pt}
\caption{Test Set Class Distribution}\label{tab2}%
\begin{tabular}{@{}ll@{}}
\toprule
Class & Count \\
\midrule
Neutrophil Segmented    & 5550 \\
Lymphocyte    & 2887 \\
Monocyte   & 1227 \\
Neutrophil Band & 825 \\
Others & 638 \\
Eosinophil & 397 \\
Basophil & 57 \\
\bottomrule
\end{tabular}
\end{minipage}
\end{table}

\subsection{Neural Color Transfer (NCT)}\label{Technique}
This research aims to investigate and subsequently utilize the color transfer techniques proposed by He et al. \cite{he2019progressive} and Wanyu et al. \cite{wanyu2020day} to meet Stain Normalization standards for hematological slides. Specifically, the goal of this study is to perform staining normalization between two blood sample images with different appearances or textures of individual WBC classes. This requires establishing a semantic correlation between the input images, which is accomplished using a VGG19 network to convert low-level features such as edges, corners, and shape information into high-level semantics such as nuclei and cytoplasm texture. To ensure suitability for this specific application scenario, the color transfer method is improved by normalizing and incorporating additional features which are outlined in section \ref{Enhancement}, rather than relying solely on the generic VGG-19 model.

\subsubsection{PatchMatch Algorithm with BDS Voting }\label{PatchMatch}
NNF (Nearest Neighbor Field) search is a technique used in computer vision and image processing to establish a correspondence between two images. Given a source image and a target image, NNF search aims to find for each pixel in the source image the nearest neighbor pixel in the target image that best matches it. The resulting nearest neighbor field can be used for various purposes, such as image completion, texture synthesis, and image style transfer. NNF search algorithms can vary in their approach and efficiency, but they all involve computing patch similarities and establishing a mapping between the source and target images based on the nearest neighbors.

\begin{algorithm}
\caption{PatchMatch Algorithm with Bidirectional Similarity Voting}
\begin{algorithmic}[1]
\State Initialize mappings randomly from S to R and from R to S
\For {$i=1$ to $num\_iterations$}
    \State Find a few random patches in both images
    \State Look for nearby patches that are also a good match
    \State Repeat the previous step a few times to find the best match
    \State Update the mappings based on the best matches found
    \State Use bidirectional similarity voting to refine mappings
\EndFor
\end{algorithmic}
\end{algorithm}

The PatchMatch \cite{barnes2009patchmatch} technique, which is a type of NNF search, is a way to find a patch in one image that is similar to a patch in another image. It's useful for things like removing objects from images or changing the size of an image. The algorithm works by looking for the best matching patches in the two images. Instead of searching the entire image for a match, the algorithm starts with a few random patches and looks for nearby patches that are also a good match. This is faster and usually finds a good match. By repeating this process a few times, the algorithm can find the best match. Bidirectional similarity voting is then used to refine the mappings. This means that the algorithm looks for matches not only from image S to image R but also from image R to image S. By doing this, the mappings become more accurate, which is especially important when dealing with complex images. The algorithm can also work on feature maps generated by a neural network.

\subsubsection{Feature Map Normalization}\label{Normalization}
Neural network layers have feature maps of varying sizes, with each channel's feature map representing a unique feature of the input image. However, specific channels may contain critical features with a narrow pixel value range that can result in a loss of accuracy during the PatchMatch process, where feature correlation occurs between the source and reference images. To address this issue, feature normalization approaches, such as standardization or normalization can be applied. This study employs standardization, which normalizes each channel's pixel values by subtracting the channel's pixel mean and dividing by its standard deviation. 

\begin{align}
 z =  \frac{x- \mu}{\omega}  \label{eq1}
\end{align}

Eq. \ref{eq1} equalizes the contribution of all features and ensures that all channels' pixel values fall within the same range, preventing high-value data from being emphasized and low-value data from being downplayed and preventing loss of accuracy during the PatchMatch process.

\subsubsection{Progressive Feature Refinement}\label{Enhancement}
To improve color transfer accuracy and establish a better correlation between the source and reference, it is effective to concatenate additional features to the original features and perform feature normalization. This approach can improve the feature vector's representational capacity. Additional features were extracted from input images using semantic segmentation and clustering. Semantic segmentation partitions an image into semantically meaningful regions by assigning a class label to each pixel based on its semantic meaning, while clustering groups similar pixels or regions based on their feature similarity. These techniques extract more discriminative features, which, when concatenated with the original features, can enhance the accuracy of the classification task.

\subsubsection{Luminance Calculation Correction}\label{Luminance}
Wanyu et. al. was using the following equation for luminance calculation:
\begin{align}
\text{luminance} = 0.2126\cdot r + 0.7152\cdot g + 0.0722\cdot b \label{eq2}
\end{align}
where $r$, $g$, and $b$ are the red, green, and blue color channel values of a given pixel. It was replaced by the following equation:
\begin{align}
\text{luminance} = 0.299\cdot r + 0.587\cdot g + 0.114\cdot b \label{eq3}
\end{align}
The two luminance equations have different coefficients for the RGB channels, which results in different luminance values for each pixel in an image.

Eq. \ref{eq2}, which is often used in computer graphics and photography, has the coefficients 0.2126, 0.7152, and 0.0722 for the red, green, and blue channels respectively. This equation is based on the fact that the human eye is more sensitive to green than to red or blue, and it also takes into account the fact that the blue channel has the least impact on the overall brightness.

Eq. \ref{eq3}, which is used in OpenCV, has the coefficients 0.299, 0.587, and 0.114 for the red, green, and blue channels respectively. These coefficients were chosen based on the CIE 1931 color space, which is a standard color space used in color science.

In general, both luminance equations provide a decent estimate of the perceived brightness of an image. However, the coefficients used in each equation may result in slightly different luminance values. The choice of which equation to use depends on the particular application and the intended outcome.

\subsubsection{Loss Function Modification}\label{Loss Function}
A common challenge in data analysis is determining the appropriate coefficients of a linear function that fits a given data distribution. This task can be accomplished by defining a loss function and adjusting it based on specific constraints.

He et al. proposed a three-part framework for constructing such an objective function. The first component, denoted as E$_D$, aims to establish a similarity between the guidance G$_L$ and the output. The second component, E$_L$, focuses on maintaining edge preservation in the source S$_L$ while ensuring similarity between locally adjacent pixels. The third component, E$_{NL}$, guarantees that pixels with similar colors in the source image will transfer to the output based on locally similar features.

By combining these three components, the overall objective function can be formulated as follows:
\begin{equation}
\text{E}_{\text{total}} = \text{E}_\text{D} + \lambda_\text{L} \ast \text{E}_\text{L} + \lambda_\text{NL} \ast \text{E}_\text{NL}
\label{eq4}
\end{equation}

In Eq.\ref{eq4}, the weights  $\lambda_\text{L}$ and  $\lambda_\text{NL}$ determine the contribution of the E$_L$ and E$_{NL}$ terms, respectively. By default, the values of these weights are set to 0.125 and 2.0, respectively. However, through trial and error, we have determined that modifying these values to 0.001 and 0.4, respectively, can allow for a more generalized color transfer that is better suited to our specific use case.

\subsubsection{Local Color Transfer}\label{LCT}
The proposed approach is based on He et al.'s color transfer algorithms and includes a suggested feature augmentation technique to improve color transfer results. However, the original implementation could only convert a light-colored image to a dark color, but we modified the architecture  by replacing the method that calculated luminance, as explained in section \ref{Luminance}, so that our desired result was achieved.

The proposed approach aims to enhance transfer speed by utilizing NNF which is typically performed on all CNN layer outputs and can be computationally expensive. To reduce the computational overhead, the proposed approach performs the NNF search and local color transfer at a single layer instead of all five layers, which can be set to a number between 1 and 5. Layer 1 provides the best performance because it has the highest image resolution.

 These techniques can help to identify and group similar pixels based on their color, texture, and other features, which can further improve the accuracy of the transfer process.

\begin{algorithm}
\caption{Neural Color Transfer \cite{he2019progressive}}\label{algo1}
\begin{algorithmic}[1]
\Require Source Image S, Reference Image R
\Ensure set L = a random number between 1 and 5
\State \textbf{Do}
    \State \hskip0.5em \textbf{NNF Search:}
    \State \hskip1em F$^L_S$, F$^L_R$ $\Leftarrow$ feed S, R to VGG19 and get features.
    \State \hskip1em \textbf{Feature Normalization:}
    \State \hskip1.5em F$^L_S$, F$^L_R$ $\Leftarrow$ normalize pixel values of F$^L_S$, F$^L_R$.
     \State \hskip1em \textbf{Feature Enhancement:}
    \State \hskip1.5em FC$^L_S$, FC$^L_R$ or FM$^L_S$, FM$^L_R$ $\Leftarrow$ apply clustering or semantic segmentation.
    \State \hskip1em $\tilde{\emptyset}^L_S{\rightarrow}_R$ $\Leftarrow$ map FC$^L_S$ to FC$^L_R$ or map FM$^L_S$ to FM$^L_R$
    \State \hskip1em $\tilde{\emptyset}^L_R{\rightarrow}_S$ $\Leftarrow$ map FC$^L_R$ to FC$^L_S$ or map FM$^L_R$ to FM$^L_S$
    \State \hskip1em G$^L$ $\Leftarrow$ reconstruct S$^L$ with R$^L$ and bidirectional mappings.
    \State \hskip0.5em \textbf{Local Color Transfer:}
    \State \hskip1em a$^L$, b$^L$ $\Leftarrow$ optimize linear transform from S$^L$, G$^L$ by minimizing the objective function \cite{wanyu2020day} using multi-label graph cut optimization 
    \State \hskip1em a$^L$, b$^L$ $\Leftarrow$ upscale a$^L$, b$^L$ with Fast Guided Filter.
    \State \hskip1em $\tilde{S}^L$ $\Leftarrow$ transfer the color of S by $\tilde{S}^L(p)= a^L_\uparrow(p) S^L (p) + b^L_\uparrow(p)$
\end{algorithmic}
\end{algorithm}

The proposed approach leverages the C language, following the precedent set by Wanyu et al., to enhance the performance of computationally intensive modules such as NNF search and local color transfer. This optimization results in a significant reduction of the overall runtime, from 5 minutes per 500x500 image in the Python implementation to just 30 seconds per 500x500 image in the C implementation.

\subsection{Evaluation Strategy}\label{Evaluation}
We conducted two experiments to evaluate the effectiveness of our stain normalization process. We trained a YOLOv5 object detector on the Lab A dataset, which served as the reference dataset. Additionally, we collected a dataset from Lab B with the same number of classes but varying appearances compared to Lab A data. We applied color balancing to both datasets and compared the detector results. Secondly, we performed NCT on both Lab A and Lab B datasets. For each dataset, we selected 1-3 representative images as reference images, and their color was transferred to the other dataset.

\begin{figure}[H]
\centering
\includegraphics[width=0.8\textwidth]{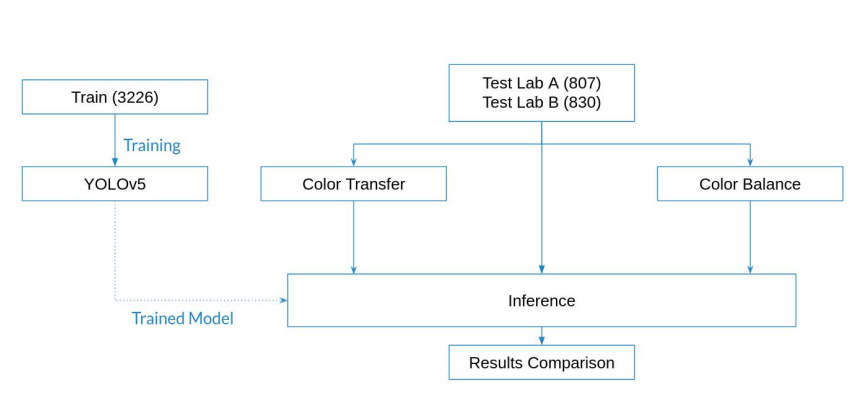}
\caption{AI-Based Evaluation Methodology}\label{fig3}
\label{fig:Eval}
\end{figure}
\begin{figure}[H]
\centering
\includegraphics[width=0.8\textwidth]{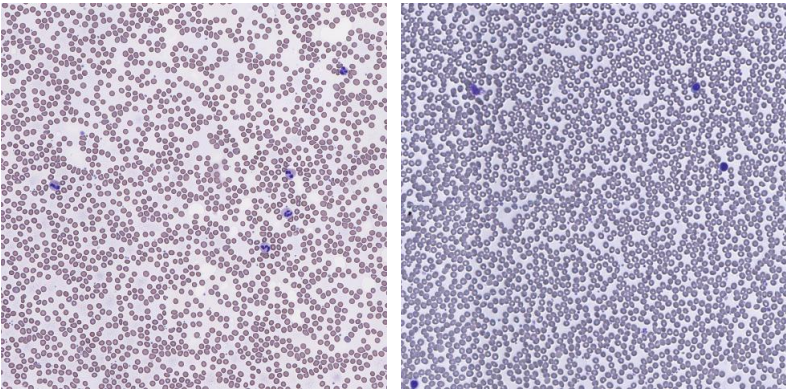}
\caption{Representative Images of Lab A and Lab B data}\label{fig4}
\label{fig:Rep}
\end{figure}

In Section \ref{Results}, we present the results of both experiments, including representative images from both labs, as shown in Figure \ref{fig:Rep}. The images from Lab A were mostly perfectly stained or over-stained, whereas those from Lab B were either under-stained or over-stained. This made the Lab B data ideal for evaluating our methodology and experiment.

\section{Results}\label{Results}
\begin{figure}[H]
\centering
\includegraphics[width=0.8\textwidth]{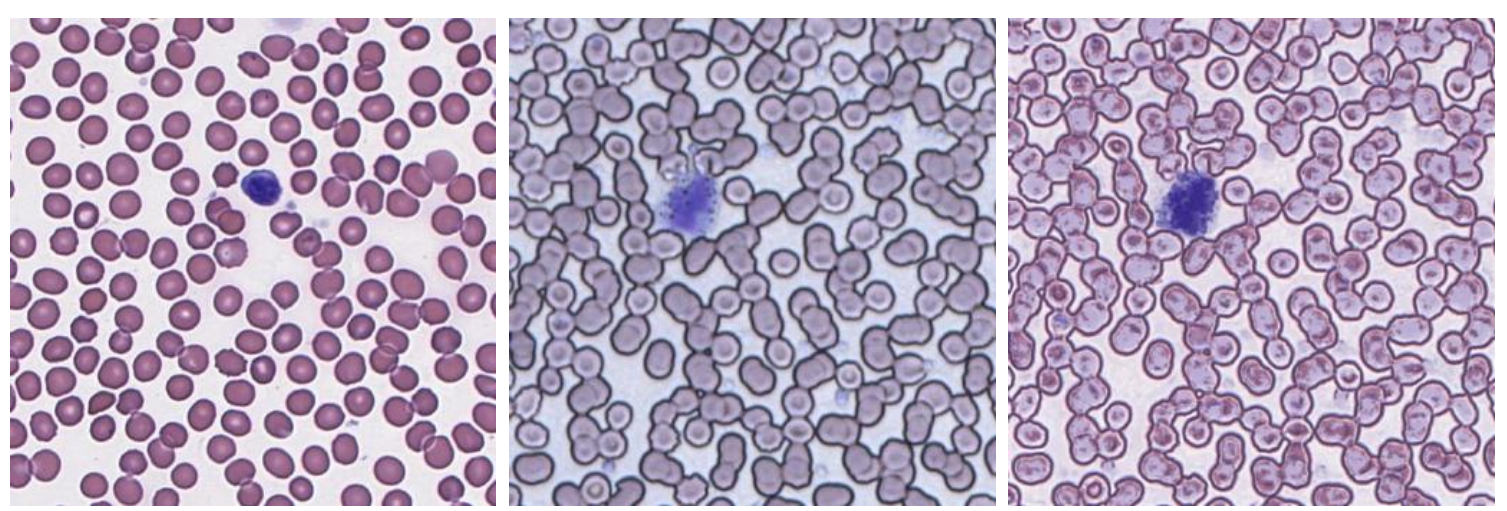}
\caption{Color Transfer from Lab A to Lab B (Feature Enhancement via Clustering)}\label{fig5}
\label{fig:A2BC}
\includegraphics[width=0.8\textwidth]{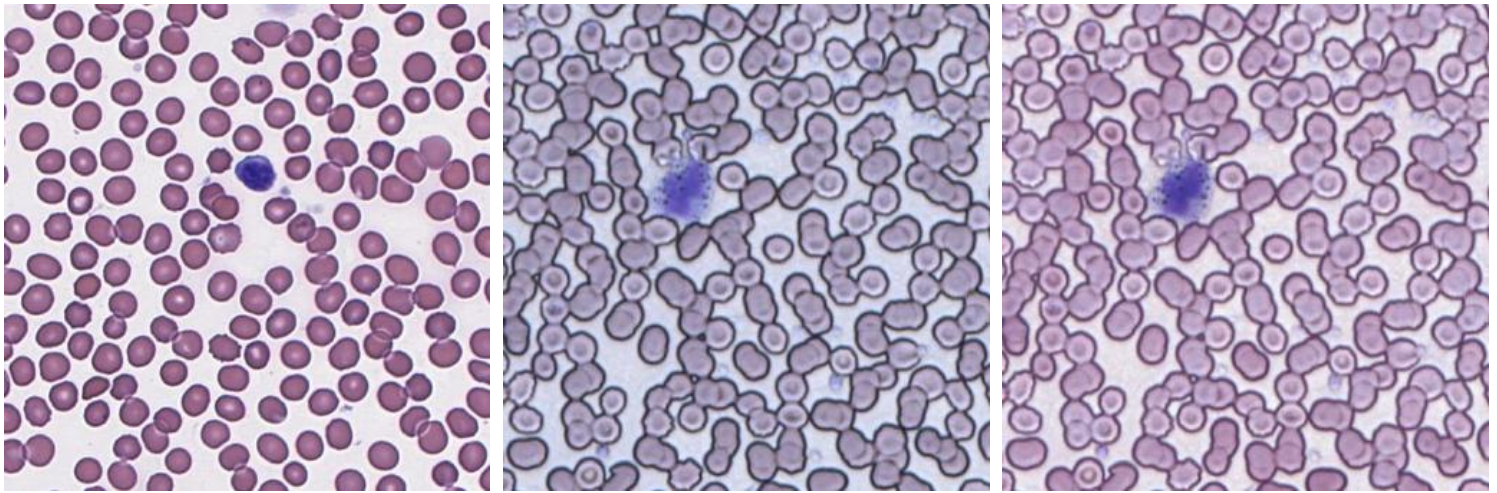}
\caption{Color Transfer from Lab A to Lab B (Feature Enhancement via Semantic Segmentation)}\label{fig6}
\label{fig:A2BS}
\end{figure}

\begin{figure}[H]
\centering
\includegraphics[width=0.8\textwidth]{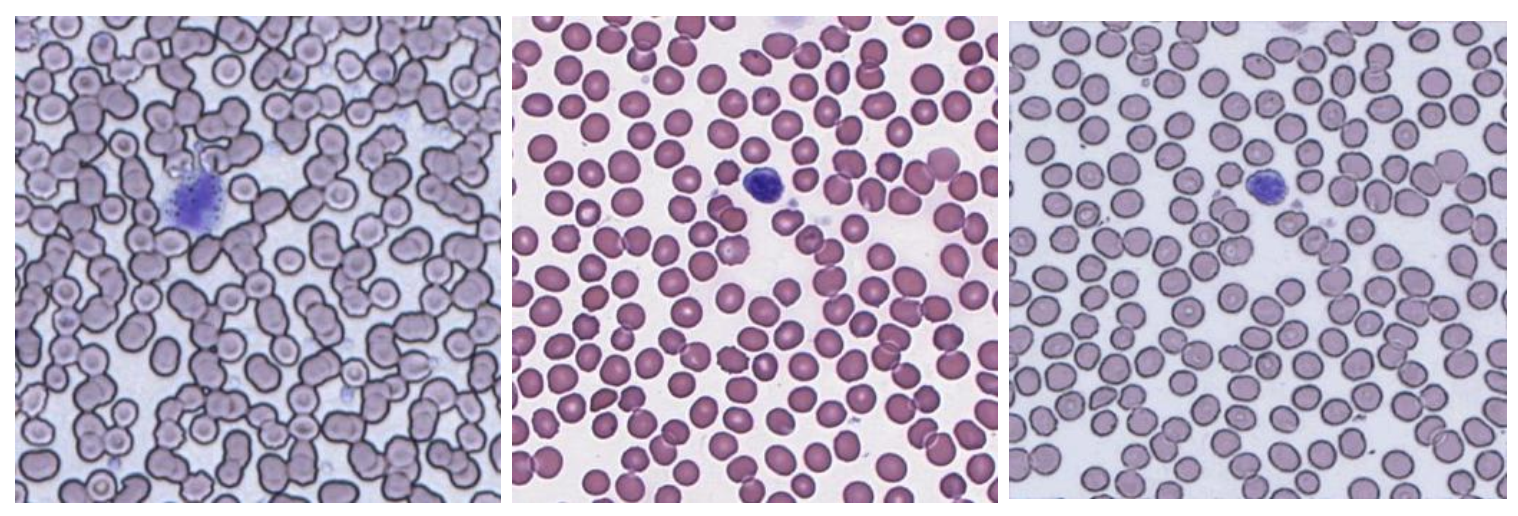}
\caption{Color Transfer from Lab B to Lab A (Feature Enhancement via Clustering)}\label{fig8}
\label{fig:B2AC}
\includegraphics[width=0.8\textwidth]{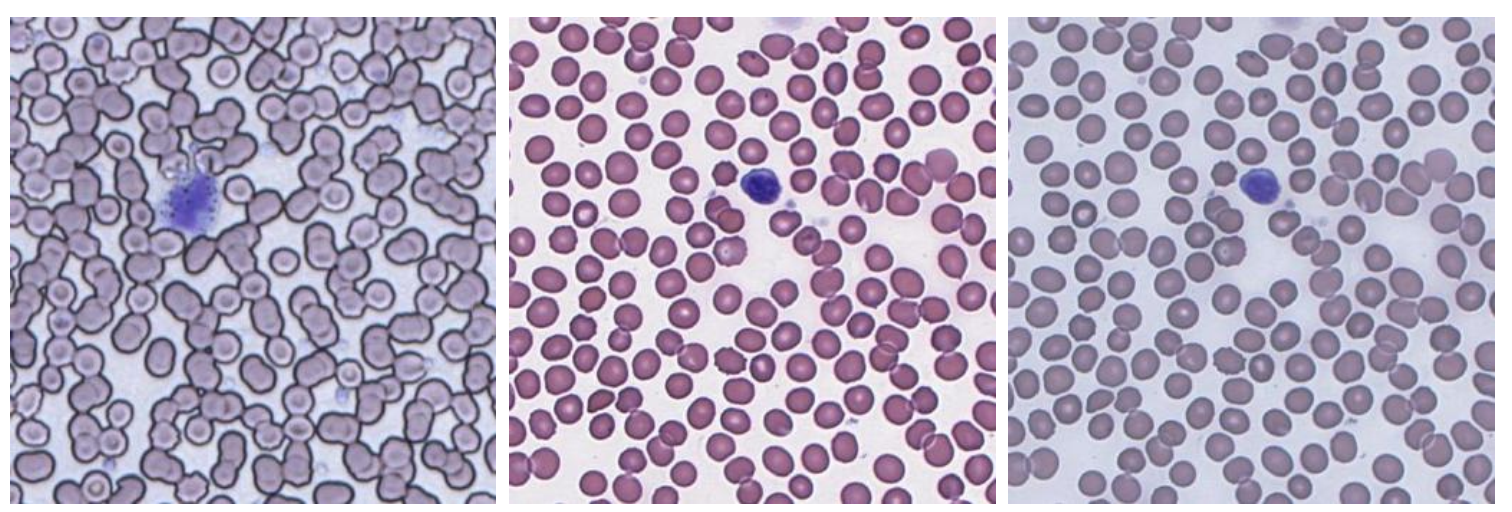}
\caption{Color Transfer from Lab B to Lab A (Feature Enhancement via Semantic Segmentation)}\label{fig9}
\label{fig:B2AS}
\end{figure}

We evaluated our stain normalization process by training a YOLOv5 object detector on our Lab A dataset, which was color-balanced along with Lab B data, and the results were compared. The results with clustering are also shown to show the difference with feature addition by semantic segmentation (Fig. \ref{fig:A2BC},\ref{fig:A2BS},\ref{fig:B2AC} and \ref{fig:B2AS}).

\begin{figure}[H]
\centering
\includegraphics[width=0.8\textwidth]{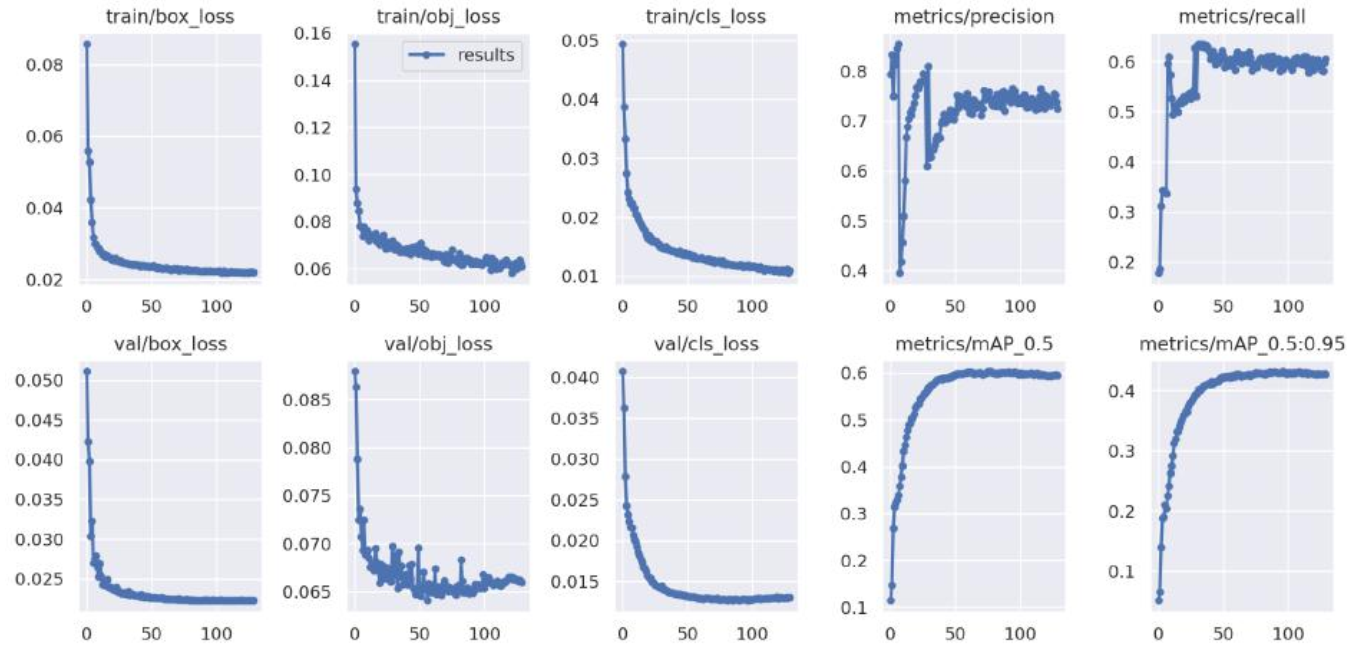}
\caption{Loss and Metric Graphs for Training and Validation of YOLOv5}\label{fig7}
\label{fig:yolomet}
\end{figure}

The YOLOv5 detector was trained on 3226 images for 150 epochs and tested on 807 images of lab A data, and the precision, recall, Pascal VOC mAP, and COCO mAP were recorded. We observed how the training and validation metrics changed throughout the training process, which is shown in Fig. \ref{fig:yolomet}.


We measured the effect of the simplest color-balancing algorithm on the object detection results and found that the performance decreased slightly. Finally, the Neural Color Transfer algorithm was applied to this dataset using a reference image from the same dataset, and the performance recorded was between the original and the color-balanced one as shown in Table \ref{LabARes}.

\begin{table}[H]
\centering
\caption{Results on Lab A data}\label{LabARes}%
\begin{tabular}{@{}llllll@{}}
\toprule
Technique & Precision  & Recall & mAP@0.5 & mAP@0.5:0.95 & Reported on\\
\midrule
Original    & 0.922   & 0.901  & 0.945  & 0.683  & Lab A Test Set  \\
Color Balance    & 0.899   & 0.885  & 0.925  & 0.65  & Lab A Test Set  \\
Color Transfer    & 0.91   & 0.891  & 0.939  & 0.674  & Lab A Test Set  \\
\bottomrule
\end{tabular}
\end{table}

On the Lab B data, we initially tested the model trained on Lab A data and recorded its metrics. After that, data B was color-balanced, and the results improved somewhat. Finally, color was transferred to this dataset using the Neural Color Transfer algorithm, and the results improved significantly, as shown in Table \ref{LabBRes}.

\begin{table}[H]
\centering
\caption{Results on Lab B data}\label{LabBRes}%
\begin{tabular}{@{}llllll@{}}
\toprule
Technique & Precision  & Recall & mAP@0.5 & mAP@0.5:0.95 & Reported on\\
\midrule
Original    & 0.499   & 0.2  & 0.346  & 0.275  & Lab B Data  \\
Color Balance    & 0.692   & 0.351  & 0.531  & 0.398  & Lab B Data \\
Color Transfer    & 0.72   & 0.405  & 0.572  & 0.467  & Lab B Data \\
\bottomrule
\end{tabular}
\end{table}

By comparing the Tables \ref{LabARes} and \ref{LabBRes}, we can see that the mAP@0.5:0.95 of the object detection model on the color-transferred data improved by 70\% on Lab B data from 0.275 to 0.467 if we view the change in unseen data's metrics with respect to Lab B data only. If we consider the relative change in accuracy with respect to Lab A data and take 0.683 (from Table \ref{LabARes}) as the maximum achievable mAP@0.5:0.95, we can say that the mAP@0.5:0.95 of our unseen data improved from 40\% to 68\% relative to 100\% accuracy of Lab A data.

\section{Conclusion}\label{Discussions}
We present a novel method in this paper to normalize semantically related hematological representations using a revised approach based on main color transfer methods that transfer both color and brightness resulting in a natural scene transfer. By adding more features and normalizing them, we have improved consistency and accuracy. The results of our study reveal that this approach is effective for stain normalization and has the potential for wider applications.

Despite our successful attempt in the field of hematology, our approach has some limitations. Firstly, the accuracy of color transfer remains a concern for computer vision researchers. In general, we need to find a better approach to construct pixel correlation between images rather than solely relying on feature maps of VGG19 to solve this issue. Although pre-trained models save time while doing semantic matching across images, they may forfeit some accuracy or adaptation. Therefore, we recommend using application-specific semantic segmentation models to improve object matching accuracy and hence color transfer accuracy. Secondly, we must carefully examine future computing costs. Presently, dealing with normal-size images takes around 60 seconds on a computer, and high-resolution images might take even longer. We suggest speeding up the process of feature mapping between images or using a faster GPU to reduce the time cost.

\bibliographystyle{unsrt}  
\bibliography{references}

\end{document}